\documentclass{article}
\usepackage{graphicx}

\begin{document}

\title{Polytropic configurations with non-zero cosmological constant}

\author{M. Merafina
\thanks{Department of Physics, University of Rome ``La Sapienza"
Piazzale Aldo Moro 2, I-00185 Rome, Italy}\,\,,
G.S. Bisnovatyi-Kogan
\thanks{Space Research Institute (IKI)
 Profsoyuznaya 84/32, Moscow 117997, Russia}\,\,,
S.O. Tarasov
\thanks{National Research Nuclear University MEPHI,
Kashirskoe Shosse 31, Moscow 115409, Russia}}

 \date{}
 \maketitle

\begin{abstract}
We solve the equation of the equilibrium of the gravitating body, with a
polytropic equation of state of the matter $P=K\rho^{\gamma}$, with
$\gamma=1+1/n$, in the frame of the Newtonian gravity, with non-zero
cosmological constant $\Lambda$. We consider the cases with
$n=1,\,\,1.5,\,\,3$ and construct series of solutions with a fixed value
of $\Lambda$. For each value of $n$, the non-dimensional equation of the
static equilibrium has a family of solutions, instead of the unique
solution of the Lane-Emden equation at $\Lambda=0$. The equilibrium state
exists only for central densities $\rho_0$ larger than the critical value
$\rho_c$. There are no static solutions at $\rho_0 < \rho_c$.
 We investigate the stability of equilibrium bodies in presence
 of nonzero $\Lambda$, and show that dark energy decease the dynamic
 stability of the configuration. We apply
our results for analyzing the properties of equilibrium
states for cluster of galaxies in the present universe with non-zero
$\Lambda$.
\end{abstract}


\section{Introduction}

Detailed analysis of the observations of distant SN Ia (Riess et
al., 1998; Perlmutter et al., 1999) and of the spectrum of
fluctuations of the cosmic microwave background radiation (CMB), see
e.g. Spergel et al. (2003), Tegmark et al. (2004), have lead to
conclusion that the term, representing ``dark energy" contains about
70\% of the average energy density in the present universe and its
properties are very close (identical) with the properties of the
Einstein cosmological $\Lambda$ term. This value of the cosmological
constant was anticipated by Kofman \& Starobinsky (1985), basing on
the analysis of the existing upper limits for the microwave
background anisotropy, see also
Eke et al. (1996). In papers of Chernin (see reviews 2001 and 2008)
the question was raised about a possible influence of the existence of the
cosmological constant on the properties of the Hubble flow in the local
galaxy cluster at close vicinity of our Galaxy. Basing on the observations
of Karachentsev et al. (2006), it was concluded by Chernin (2001, 2008)
that the presence of the the dark energy (DE) is responsible for the
formation of this Hubble flow. This conclusion was criticized by
Lukash \& Rubakov (2008).

The importance of the DE for the structure of the local galaxy
cluster (LC) depends on the level of the influence of DE on the
dynamic properties. In particular, it is necessary to check, if the
LC may exist in the equilibrium state, at present values of DE
density, and the LC densities of matter, consisting of the baryonic
and dark matter (BM and DM).

Here, we construct, in Newtonian approximation, exact equilibrium
configurations of the selfgravitating objects in presence of DE. We
consider a polytropic equation of state, having in mind, that very
extended configurations, like LC,
 where the dark matter have not reached the state of a full
thermodynamic equilibrium, and effective temperature could be not a
constant over the cluster. In presence of DE ($\Lambda$) the
polytropic configurations are represented by a family of
configurations, instead of the single model for each polytropic index
$n$. The additional parameter $\beta$ represents the ratio of the density
of DE to the matter central density of the configuration. For values
of $n=1,\,\,1.5,\,\,3$, corresponding to the polytropic powers
$\gamma=2,\,\,5/3,\,\,4/3$, we find the limiting values of $\beta_c$, so
that at $\beta>\beta_c$ there are no equilibrium configurations and
only expanding cluster, may be with a Hubble flow, may exist. We
apply these results to the LC using the observational data and
different suggestions, connected with DM distribution. Note that the
structure of the isothermal configurations in a box, in presence of
DE was investigated by de Vega \& Siebert (2005): in this paper the
authors were interested in the problem of thermodynamic stability of
such configurations, following the works of Antonov (1962) and
Lynden-Bell \& Wood (1968).
We have derived a virial theorem for the equilibrium configurations
in presence of DE, and investigate an influence of the DE on the
dynamic stability of the equilibrium configurations, using an
approximate energetic method. It is shown that DE produce a
destabilizing effect, contrary to the stabilizing  influence of the
cold dark matter (McLaughlin \& Fuller, 1996).

\section{Main equations}

Let us consider spherically symmetric equilibrium configuration in
Newtonian gravity, in presence of DE, represented by the cosmological
constant $\Lambda$. In this case, the gravitational force $F_g$ which a
unit mass undergoes in a spherically symmetric body is given by the
formula (Chandrasekhar, 1939)

\begin{equation}
F_g=-\frac{Gm}{r^2}+\frac{\Lambda r}{3},
 \label{eq1}
\end{equation}
where $m$ is the mass of the matter inside the radius $r$, which is
connected with the matter density $\rho$ by the continuity equation

\begin{equation}
 \label{eq2}
 \frac{dm}{dr}=4\pi\rho r^2.
\end{equation}
The density of DE $\rho_v$ is connected with $\Lambda$ as

\begin{equation}
\rho_v=\frac{\Lambda}{8\pi G}.
 \label{eq3}
\end{equation}
Then, the equilibrium equation of the selfgravitating body is given by

\begin{equation}
 \frac{1}{\rho}\frac{dP}{dr}=-\frac{Gm}{r^2}+\frac{\Lambda r}{3}.
 \label{eq4}
\end{equation}
Now, let us consider a polytropic equation of state

\begin{equation}
 P=K\rho^{\gamma},\quad {\rm with}\quad \gamma=1+\frac{1}{n}.
 \label{eq5}
\end{equation}
By introducing the nondimensional polytropic variables $\xi$ and
$\theta_n$ so that (Chandrasekhar, 1939)

\begin{equation}
 r=\alpha\xi \quad {\rm and}\quad \rho=\rho_0 \theta_n^n,
 \label{eq6}
\end{equation}
we obtain, from (\ref{eq2}) and (\ref{eq4}), the equation describing the
structure of the polytropic configuration in presence of the DE as

\begin{equation}
 \frac{1}{\xi^2}\frac{d}{d\xi}\left(\xi^2\frac{d\theta_n}{d\xi}\right)=
 -\theta_n^n+\beta.
 \label{eq7}
\end{equation}
Here $P$ is the pressure and $\rho_0$ is the matter central density,
$\alpha$ is the characteristic radius given by

\begin{equation}
\alpha^2=\frac{(n+1)K}{4\pi G}\rho_0^{\frac{1}{n}-1},
 \label{eq6a}
\end{equation}
while $\beta={\Lambda}/{4\pi G\rho_0}={2\rho_v}/{\rho_0}$
represents the ratio of the DE density to the central density of the
configuration. At $\beta=0$ we obtain from (\ref{eq7}) the well known
Lane-Emden equation (Chandrasekhar, 1939).

Note, that the generalization of Lane-Emden equation in presence of $\Lambda$ was first obtained by Balaguera-Antol\'inez et al. (2007).

\section{The virial theorem}

Let us find the relations between gravitational $\varepsilon_g$ and
thermal $\varepsilon_{th}$ energies, and the energy
$\varepsilon_\Lambda$, of the interaction of the body with DE. The
gravitational energy of the selfgravitation is written as (Landau
\& Lifshitz, 1980)

\begin{equation}
 \varepsilon_g=-G\int_0^{M} \frac{m dm}{r},
\quad m=4\pi\int_0^r\rho {r'}^2 dr'
=-\frac{r^2}{G}\left(\frac{1}{\rho}\frac{dP}{dr}-\frac{\Lambda
r}{3}\right),\quad M=m(R).
 \label{v1}
\end{equation}
Here $m$ is written using the equilibrium equation (\ref{eq4}). For
the adiabatic stars with a polytropic equation of state we have the
relations

\begin{equation}
\rho E=nP,\quad I=E+\frac{P}{\rho}=\frac{n+1}{n}E,
 \label{v2}
\end{equation}
where $E$ and $I$ are thermal energy and enthalpy per mass unit. After
some transformations, similar to ones made by Landau \& Lifshitz (1980),
we obtain the following relations between the energies

\begin{equation}
 \varepsilon_g=-\frac{3}{n} \varepsilon_{th}+2 \varepsilon_\Lambda,
 \qquad
\varepsilon_{th}=-\frac{n}{3} \varepsilon_{g}+\frac{2n}{3} \varepsilon_\Lambda,
 \label{v4}
\end{equation}

\begin{equation}
 \varepsilon_{tot}=\varepsilon_{th}+\varepsilon_{g}+\varepsilon_\Lambda=
 \frac{3-n}{3} \varepsilon_{g}+\frac{2n+3}{3} \varepsilon_\Lambda=
 \frac{n-3}{n} \varepsilon_{th}+3\varepsilon_\Lambda.
 \label{v5}
\end{equation}
where $\varepsilon_\Lambda=-\frac{4\pi G\rho_v}{3}\int_0^M r^2
dm=-\frac{\Lambda}{6}\int_0^M r^2 dm$ is defined below in
(\ref{eq27}) and the additive constant in the energy definition
of $\varepsilon_\Lambda$ is chosen so that $\varepsilon_\Lambda=0$
at $\Lambda=0$ and $M=0$, being $\varepsilon_{th}=\int_0^M E\,dm$. Now,
let us make an additional transformation of the expression for
$\varepsilon_g$. Then, by partial integration, we obtain

\begin{equation}
 \varepsilon_g=-G\int_0^{M} \frac{m dm}{r}=
 -\frac{GM^2}{2R}-\frac{G}{2}\int_0^{R} \frac{m^2}{r^2}dr.
 \label{v6}
\end{equation}
We can transform the last integral by using (\ref{v1}), (\ref{v2}) and
making partial integrations. We have

$$ \frac{G}{2}\int_0^{R} \frac{m^2}{r^2}dr=-\frac{1}{2}\int_0^R
 r^2\left(\frac{1}{\rho}\frac{dP}{dr}-\frac{\Lambda
r}{3}\right)\frac{m}{r^2}
dr=-\frac{1}{2}\int_0^R\frac{m}{\rho}\frac{dP}{dr}dr+\frac{\Lambda}{6}\int_0^R
mr\,dr
$$
\begin{equation}
=\frac{1}{2}\int_0^M I\,dm + \frac{\Lambda}{12}MR^2- \frac{\Lambda}{12}\int_0^M
 r^2\,dm=\frac{n+1}{2n} \varepsilon_{th} + \frac{\Lambda}{12}MR^2+
 \frac{\varepsilon_\Lambda}{2}.
 \label{v7}
\end{equation}
Then, by using (\ref{v4}), (\ref{v7}), we obtain from (\ref{v6}) the
following relations

\begin{equation}
 \varepsilon_g=-\frac{3}{5-n}\frac{GM^2}{R}-\frac{\Lambda}{2(5-n)}MR^2
 -\frac{2n+5}{5-n}\varepsilon_\Lambda,
 \label{v8}
\end{equation}

\begin{equation}
 \varepsilon_{th}=\frac{n}{5-n}\frac{GM^2}{R}+\frac{n\Lambda}{6(5-n)}MR^2
 +\frac{5n}{5-n}\varepsilon_\Lambda.
 \label{v9}
\end{equation}
Finally, by inserting (\ref{v9}) in (\ref{v5}), we get

\begin{equation}
\varepsilon_{tot}=\frac{n-3}{5-n}\frac{GM^2}{R}+\frac{(n-3)\Lambda}{6(5-n)}MR^2
 +\frac{2n}{5-n}\varepsilon_\Lambda.
 \label{v10}
\end{equation}
We can calculate $\varepsilon_{tot}$ for some particular cases. For
$n=3,\,\,1,\,\, {\rm and}\,\, 0$ we have, respectively.

\begin{equation}
 \varepsilon_{tot}=3\varepsilon_\Lambda,\quad \varepsilon_{tot}=-\frac{1}{2}
\frac{GM^2}{R}-\frac{1}{12}\Lambda MR^2+\frac{1}{2}\varepsilon_\Lambda,
\quad \varepsilon_{tot}=-\frac{3}{5} \frac{GM^2}{R}-\frac{1}{10}\Lambda
MR^2,
 \label{v11}
\end{equation}
The Lane-Emden model with $n=5$, has an analytical solution with finite
mass $M$, finite values of the energies and infinite radius $R$, so that
must be $(5-n)R\, \rightarrow \,{\rm const}$, if $n\rightarrow 5$. In
presence of DE the finiteness of values of all types of energies require
that

\begin{equation}
 (5-n)R\, \rightarrow \, {\rm const} \quad {\rm and} \quad
 \varepsilon_\Lambda\, \rightarrow \, -\frac{\Lambda}{30} MR^2,\,\,
{\rm if}\,\,\, n\rightarrow 5.
 \label{v12}
\end{equation}
The Lane-Emden solution (without DE) at $n=3$ has the zero total energy,
with a given finite radius and corresponds to a neutral equilibrium. So,
the knowledge of the total energy of the configuration permits to find the
boundary between dynamically stable ($n<3$, $\varepsilon_{tot}<0$) and
unstable ($n>3$, $\varepsilon_{tot}>0$) configurations. In our case the
virial theorem does not permit to do it, because the value of
$\varepsilon_\Lambda$ is not properly defined, while the presence of DE in
the whole space does not permit to chose, in a simple way, a universal
additive constant of the energy. Therefore, in spite of the result from
(\ref{v11}), where $\varepsilon_{tot}=3\varepsilon_\Lambda<0$ at $n=3$, in
accordance to the stability analysis made in section 4.3, the polytropic
solution at $n=3$ in presence of DE becomes unstable.

The virial theorem in presence of $\Lambda$ was investigated by Balaguera-Antol\'inez et al. (2007). The final result in this paper was presented in the form of integrals, instead of simple algebraic relations obtained above.

\section{Equilibrium solutions}

\subsection{The case n=1}

At $n=1$ the equation (\ref{eq7}) is linear and has an analytic
solution. Following Chandrasekhar (1939), let us introduce a new
variable $\chi=\xi\theta_1$. After the substitution the equation
(\ref{eq7}) is transformed into

\begin{equation}
 \frac{d^2\chi}{d\xi^2}=-\chi+\xi\beta.
 \label{eq8}
\end{equation}
At $\beta=0$ we have a well known analytic polytropic solution
(Chandrasekhar, 1939)

\begin{equation}
 \theta_1=\frac{\sin\xi}{\xi},
 \label{eq9}
\end{equation}
satisfying the boundary conditions in the center $\theta_1(0)=1$,
$\theta_1'(0)=0$. At nonzero $\beta$, the solution satisfying the
boundary conditions in the center is written as

\begin{equation}
 \theta_1=(1-\beta)\frac{\sin\xi}{\xi} +\beta.
 \label{eq10}
\end{equation}
The radius of the configuration is defined by the first zero of the
solution (\ref{eq9}), or (\ref{eq10}). For the Lane-Emden solution
(\ref{eq9}) the outer radius corresponds to $\xi_{out}=\pi$ and, for
the solution of (\ref{eq10}) with DE, the radius of the
configuration is determined by the transcendental equation

\begin{equation}
 (1-\beta)\frac{\sin\xi_{out}}{\xi_{out}}+\beta=0.
 \label{eq11}
\end{equation}

This equation has real solutions only at $\beta<\beta_c$, so that at the
outer boundary not only $\theta_1=0$, but also $\theta_1'=0$ for
$\beta=\beta_c$. It follows from (\ref{eq10})

\begin{equation}
 \theta_1'=(1-\beta)\left(\frac{\cos\xi}{\xi}-\frac{\sin\xi}{\xi^2}\right).
 \label{eq11a}
\end{equation}
therefore the parameters $\beta_c$ and $\xi_{out,c}$ of the limiting
equilibrium solution in presence of DE are determined by the algebraic
equations

\begin{equation}
(1-\beta_c)\frac{\sin\xi_{out,c}}{\xi_{out,c}}+\beta_c=0,\quad
\tan\xi_{out,c}=\xi_{out,c},\quad
\pi\,<\,\xi_{out,c}\,<\,\frac{3\pi}{2}.
 \label{eq12}
\end{equation}
The density distribution for the equilibrium configurations with values
$\beta=0,\,\,\beta=0.5\beta_c,\,\,\beta=\beta_c$ is shown in Fig.1. The
nonphysical solution at $\beta=1.5\beta_c$, which has not an outer
boundary, is given by the dash-dot line. The nonphysical parts of the
solutions at $\beta\le\beta_c$, behind the outer boundary, are given by
dash lines. At large $\xi$, these solutions asymptotically approach the
horizontal line $\theta_1=\beta$. Numerical solutions of equations
(\ref{eq12}) and (\ref{eq11}) give $\beta_c=0.178$ and $\xi_{out}=\pi,\,\,
3.490,\,\,4.493$, for $\beta=0,\,\,\beta=0.5\beta_c=0.089,\,\,\beta=
\beta_c=0.178$, respectively.

\subsection{Some general relations}

The equilibrium mass $M_n$ for a generic polytropic configuration which is
solution of the Lane-Emden equation is written as

\begin{equation}
M_n=4\pi \int_0^{R}\rho r^2 dr=4\pi \left[\frac{(n+1)K}{4\pi
G}\right]^{3/2}\rho_0^{\frac{3}{2n}-\frac{1}{2}}
\int_0^{\xi_{out}}\theta_n^n\xi^2 d\xi.
 \label{eq13}
\end{equation}
Using equation (\ref{eq7}), the integral in the right site may be
calculated by partial integration, what gives the following relation
for the mass of the configuration

\begin{equation}
M_n=4\pi \left[\frac{(n+1)K}{4\pi
G}\right]^{3/2}\rho_0^{\frac{3}{2n}-\frac{1}{2}}
\left[-\xi_{out}^2\left(\frac{d\theta_n}{d\xi}\right)_{out}+
\frac{\beta\xi_{out}^3}{3}\right].
 \label{eq14}
\end{equation}
Reminding (\ref{eq6a}), we can write the expression for the mass in
the form

\begin{equation}
M_n=4\pi \rho_0
\alpha^3\left[-\xi_{out}^2\left(\frac{d\theta_n}{d\xi}\right)_{out}+
\frac{\beta\xi_{out}^3}{3}\right].
 \label{eq15}
\end{equation}
Let us note, that here $\theta_n(\xi)$ is not a unique function, but
it depends on the parameter $\beta$, according to (\ref{eq7}). For
the limiting configuration, with $\beta=\beta_c$, we have on the
outer boundary $\xi_{out}$

\begin{equation}
\theta_n(\xi_{out})=0,\quad \frac{d\theta_n}{d\xi}|_{\xi_{out}}=0.
 \label{eq17}
\end{equation}
Therefore the mass $M_{n,lim}$ of the limiting configuration, with
account of (\ref{eq6}), is written as

\begin{equation}
M_{n,lim}=4\pi \rho_{0c} \alpha^3
\frac{\beta_c\xi_{out}^3}{3}=\frac{4\pi}{3}r_{out}{^3}\beta_c\rho_{0c}=
\frac{4\pi}{3}r_{out}{^3}\bar\rho_{c},
 \label{eq18}
\end{equation}
so that the limiting value $\beta_c$ is exactly equal to the ratio
of the average matter density $\bar\rho_{c}$ of the limiting
configuration, to its central density $\rho_{0c}$

\begin{equation}
\beta_c=\frac{\bar\rho_{c}}{\rho_{0c}}.
 \label{eq19}
\end{equation}
For the Lane-Emden solution (with $\beta=0$) we have
$\rho_{0}/\bar\rho = 3.290,\,\,5.99,\,\,54.18$ for $n=1,\,\,1.5,\,\,3$,
respectively. Let us consider the curve $M(\rho_0)$ for a constant DE
density $\rho_v=\Lambda/8\pi G$. For plotting this curve in the
nondimensional form, we introduce the characteristic density $\rho_{ch}$
and write the expression for the mass in the form

\begin{equation}
M_n=4\pi \left[\frac{(n+1)K}{4\pi
G}\right]^{3/2}\rho_{ch}^{\frac{3}{2n}-\frac{1}{2}}
\hat\rho^{\frac{3}{2n}-\frac{1}{2}}
\left[-\xi_{out}^2\left(\frac{d\theta_n}{d\xi}\right)_{out}+
\frac{\beta\xi_{out}^3}{3}\right],
 \label{eq20}
\end{equation}
where $\hat\rho_0=\rho_0/\rho_{ch}$ is the nondimensional central density
of the configuration. We introduce also the nondimensional mass of the
confuguration ${\hat M_n}$ as

\begin{equation}
M_n=4\pi \left[\frac{(n+1)K}{4\pi
G}\right]^{3/2}\rho_{ch}^{\frac{3}{2n}-\frac{1}{2}}{\hat M_n},\quad
{\hat M_n}=\hat\rho_0^{\frac{3}{2n}-\frac{1}{2}}
\left[-\xi_{out}^2\left(\frac{d\theta_n}{d\xi}\right)_{out}+
\frac{\beta\xi_{out}^3}{3}\right].
 \label{eq21}
\end{equation}
Now we can plot the nondimensional curve ${\hat M_n}({\hat\rho_0})$,
at constant $\rho_v=\beta\rho_0/2$ (from the definition of $\beta$). It is
convenient to construct the curve starting from the model with
$\hat\rho_0=1$ and $\beta\ll 1$, very close to the Lane-Emden solution,
and than following the sequence by decreasing the central density
$\hat\rho_0$ with taking $\beta\propto 1/\hat\rho_0$ (being constant
$\rho_v$) until the value $\beta=\beta_c$. For $n=1$, using (\ref{eq11a}),
we have

\begin{equation}
 \hat M_1=\hat\rho_0\left[(1-\beta)(\sin\xi_{out}-\xi_{out}\cos\xi_{out})
+ \frac{\beta \xi_{out}^3}{3}\right],\quad
 \hat\rho_0\beta=\beta_{in}\,=\,{\rm const}.
 \label{eq22}
\end{equation}
The behavior of $\hat M_1(\hat\rho_0)|_\Lambda$ (at constant $\rho_v$) is
given in Fig.2 for different values of
$\beta_{in}=0,\,\,\beta_{in}=0.5\beta_{c},\,\,\beta_{in}=\beta_{c}$, for
which $\hat M_1=\pi,\,\,3.941,\,\,5.397$ at $\hat\rho_0=1$.
We note that for $\beta_{in}=\beta_{c}$ there are equilibrium models
only for $\hat\rho_0>1$.

\subsection{The case n=3. Dynamic stability}

The equilibrium equation for this case is

\begin{equation}
 \frac{1}{\xi^2}\frac{d}{d\xi}\left(\xi^2\frac{d\theta_3}{d\xi}\right)=-\theta_
3^3+\beta.
 \label{eq23}
\end{equation}
The mass of the configuration is given by

\begin{equation}
 M_3=4\pi \left[\frac{K}{\pi
G}\right]^{3/2}{\hat M_3},\quad {\hat M_3}=
-\xi_{out}^2\left(\frac{d\theta_3}{d\xi}\right)_{out}+
\frac{\beta\xi_{out}^3}{3}.
 \label{eq24}
\end{equation}

The Lane-Emden model ($\beta=0$) has a unique value of the mass,
independent on the density (equilibrium configuration with neutral dynamical
stability). At $\beta\neq 0$ the dependence on the density appears because
the function $\theta_3$ is different for different values of $\beta$ and,
along the curve $\hat M_3(\hat\rho_0)|_\Lambda$, the value of $\beta$ is
inversely proportional to $\hat\rho_0$ (like in the case $n=1$).

The density distribution for equilibrium configurations with
$\beta=0,\,\,\beta=0.5\beta_c,\,\,\beta=\beta_c$ is shown in Fig.3. The
nonphysical solution at $\beta=1.5\beta_c$, which has not an
outer boundary, is given by the dash-dot line. The nonphysical parts of
the solutions at $\beta\le \beta_c$, behind the outer boundary, are
given by the dash lines. At large $\xi$, these solutions asymptotically
approach the horizontal line $\theta_3=\beta^{1/3}$, with damping
oscillations around this value. The numerical solution of the equation
(\ref{eq23}) gives $\beta_c=0.006$, $\xi_{out}=6.897,\,\,7.489,\,\,9.889$,
for $\beta=0,\,\,\beta=0.5\beta_c=0.003,\,\,\beta=\beta_c=0.006$, respectively.
In Fig.4 we show the behavior of $\hat M_3(\hat\rho_0)|_\Lambda$, for
different values
of $\beta_{in}=0,\,\,\beta_{in}=0.5\beta_{c},\,\,\beta_{in}=\beta_{c}$,
for which $\hat M_3=2.018,\,\,2.060,\,\,2.109$, at $\hat\rho_0=1$,
respectively.

The behavior of $\hat M_3(\hat\rho_0)|_\Lambda$ in Fig.4, showing
a decreasing mass with the increase of the central density,
corresponds, for an adiabatic index equal to the polytropic one, to
dynamically unstable configurations, according to the static
criterium of stability (Zel'dovich, 1963). When the vacuum influence
is small, it is possible to investigate the stability of the
adiabatic configuration by the approximate energetic method
(Zeldovich \& Novikov, 1966; Bisnovatyi-Kogan, 2001). For $n=3$, at $\rho_0 \gg \rho_v$, the
density in the configuration is distributed according to the Lane-Emden
solution $\rho=\rho_0\, \theta_3^3(\xi)$. In this case we may
investigate the stability to homologous perturbations, changing only
the central density at fixed density distribution, given by the
function $\theta_3$. Let us first calculate the newtonian
gravitational energy of the configuration at presence of the
cosmological constant (DE). For spherical configurations, the Poisson
equation for the gravitational potential $\varphi_*$, in presence of
DE is given by

\begin{equation}
\Delta\varphi_*=\frac{1}{r^2}\frac{d}{dr}\left(r^2\frac{d\varphi_*}{dr}
\right)=4\pi G(\rho-2\rho_v), \quad \varphi_*=\varphi+\varphi_\Lambda.
 \label{eq25}
\end{equation}

The gravitational energy of a spherical body $\varepsilon_g$,
connected with its selfgravity, is written as (Landau \& Lifshitz, 1980)

\begin{equation}
 \varepsilon_g=-G\int_0^{M} \frac{m dm}{r},
\quad m=4\pi\int_0^r\rho r^2 dr,\quad M=m(R),
 \label{eq26}
\end{equation}
where $R$ is the radius of body. Here, differently from the gravitational
potential $\varphi$ where we have the normalization $\varphi =0$ at
$r=\infty$, for $\varphi_\Lambda$ with uniform density $\rho_v$ this
normalization is not possible. Then we can choose $\varphi_\Lambda =0$
at $r=0$ as the most convenient normalization. This choice, together
using the equation (\ref{eq25}), leads to the following expression

\begin{equation}
\varphi_\Lambda=-\frac{4\pi G\rho_v}{3}r^2.
 \label{eq27a}
\end{equation}

Consequently, the gravitational energy $\varepsilon_\Lambda$,
connected with the interaction of the matter with DE, is given by

\begin{equation}
\varepsilon_\Lambda=\int_0^M \varphi_\Lambda dm=-\frac{4\pi
G\rho_v}{3}\int_0^M r^2 dm.
 \label{eq27}
\end{equation}

In order to analyze the effects due to the presence of DE on the dynamical
stability of spherical systems with a polytropic equation of state, it is
necessary to consider configurations close to the polytropic (adiabatic)
equilibrium solution at $n=3$ (and $\beta =0$), where the turning point
of stability is expected. In this case, the presence of DE does not affect
significantly the gravitational equilibrium because the nondimensional
density profile of the configuration keeps constant during the adiabatic
perturbation (homologous displacement), so that we can neglect the term
depending on $\Lambda$ in the Lane-Emden equation and the nondimensional
quantities $\xi_{out}$ and $\theta_3(\xi)$ of the unperturbed polytropic
solution at $n=3$ can be used in calculating the gravitational energy
$\varepsilon_g$ connected to the selgravity. Therefore, also the terms
depending on $\Lambda$ which appear in the expression (\ref{v8}) of
$\varepsilon_g$ can be deleted and the gravitational energy
$\varepsilon_*$ becomes

\begin{equation}
\varepsilon_*=\varepsilon_g+\varepsilon_\Lambda=-G\int_0^{M}\frac{m
\,dm}{r}-\frac{4\pi G\rho_v}{3}\int_0^M r^2 dm= -\frac{3}{2}\frac{GM^2}{R}
- \frac{\Lambda}{6}\int_0^M r^2 dm,
 \label{eq28}
\end{equation}
where, taking into account the nondimensional variables (\ref{eq6}), the
energy $\varepsilon_\Lambda$ can be written as

\begin{equation}
 \varepsilon_\Lambda= -\frac{\Lambda}{6}\int_0^M r^2 dm= -\frac{2}{3}\pi
\rho_0\alpha^5\Lambda \int_0^{\xi_{out}}\theta_3^3\xi^4 d\xi.
 \label{eq29}
\end{equation}
Here, the last integral in (\ref{eq29}) has been calculated by
Bisnovatyi-Kogan (2001) so that

\begin{equation}
 \int_0^{\xi_{out}}\theta_3^3\xi^4 d\xi=10.85.
 \label{eq30}
\end{equation}
In the analysis of the dynamical stability, let us consider the total
energy $\varepsilon$ of the configuration. Then, in addition to the
Newtonian gravitational energy $\varepsilon_g$ and the energy
$\varepsilon_\Lambda$, we must also take into account the contribution of
the thermal energy $\varepsilon_{th}$, corresponding to a specific energy
density $E$ (energy per unit mass), and we may include a small correction
$\varepsilon_{GR}$ due to general relativity (Bisnovatyi-Kogan, 2001).
We obtain

\begin{equation}
 \varepsilon=\varepsilon_{th}+\varepsilon_{g}+\varepsilon_\Lambda+
\varepsilon_{GR}=
\label{eq31}
\end{equation}
$$
 =\int_0^M E\,dm-0.639GM^{5/3}\rho_0^{1/3}-0.104\Lambda
 M^{5/3}\rho_0^{-2/3} - 0.918\frac{G^2 M^{7/3}}{c^2} \rho_0^{2/3},
$$
where we used the following relations for the polytropic configuration
with $n=3$:
\begin{equation}
 \xi_{out}=6.897 \quad {\rm and} \quad R=\alpha\xi_{out}
 =\frac{M^{1/3}\rho_0^{-1/3}}{0.426}.
 \label{eq32}
\end{equation}
The equilibrium configuration is determined by the zero of the first
derivative of $\varepsilon$ over $\rho_0$, at constant entropy $S$ and
mass $M$, while the stability of the configuration is analyzed through the
sign of the second derivative: if positive, the configuration is
dynamically stable, if negative, the configuration is unstable.

It is more convenient to take derivatives over $\rho_0^{1/3}$ than over
$\rho_0$. Then, we have

\begin{equation}
\frac{\partial\varepsilon}{\partial\rho_0^{1/3}}=
3\rho_0^{-4/3}\int_0^M P\frac{dm}{\phi(m/M)}-0.639GM^{5/3}+0.208\Lambda
 M^{5/3}\rho_0^{-1} - 1.84\frac{G^2 M^{7/3}}{c^2} \rho_0^{1/3}=0
 \label{eq33}
\end{equation}
for the equilibrium configuration and

\begin{equation}
\frac{\partial ^2\varepsilon}{\left({\partial\rho_0^{1/3}}\right)^2}=
 9\rho_0^{-5/3}\int_0^M\left(\gamma-\frac{4}{3}\right) P\frac{dm}{\phi(m/M)}
-0.623 \Lambda M^{5/3}\rho_0^{-4/3} - 1.84\frac{G^2 M^{7/3}}{c^2}
 \label{eq34}
\end{equation}
for the analysis of the dynamical stability. Here

\begin{equation}
 \gamma=\left(\frac{\rho}{P}\frac{\partial P}{\partial\rho}
 \right)_S \quad {\rm and}\quad \rho=\rho_0\phi\left(\frac{m}{M}\right)
 \label{eq35}
\end{equation}
are the adiabatic index $\gamma$ at constant entropy $S$ and the
nondimensional function $\phi$, remaining constant during homologous
perturbations, respectively. It follows from (\ref{eq34}) that DE input in
the stability of the configuration is negative and similar to the
influence of the general relativistic correction (Chandrasekhar, 1964;
Merafina \& Ruffini, 1989). On the contrary, the presence of the dark
matter, as a background, has a positive influence, increasing the dynamic
stability of the configuration (McLaughlin \& Fuller, 1996;
Bisnovatyi-Kogan, 1998). Therefore, an adiabatic star with a polytropic
index $4/3$ which is in the state of a neutral stability in the pure
Newtonian gravity, becomes unstable in presence of DE.

The presence of DE decrease the the level of relativity of a warm
dark matter (WDM), at which it is still possible to form gravitationally
bound configurations from WDM.

Note that dynamic stability of pure polytropic models was investigated by Balaguera-Antol\'inez et al. (2007), using static criterium of stability based on $M(\rho_c)$ dependence (\ref{eq20}). Our criterium (\ref{eq34}) is valid for any equation of state $P(\rho)$, for configurations near the point of the loss of stability. Influence of DE on the position of the bifurcation point for ellipsoid-spheroid transition was investigated by Balaguera-Antol\'inez et al. (2006).

\subsection{The case n=1.5}

The equilibrium equation for this case is

\begin{equation}
 \frac{1}{\xi^2}\frac{d}{d\xi}\left(\xi^2\frac{d\theta_{3/2}}{d\xi}\right)
 =-\theta_{3/2}^{3/2}+\beta.
 \label{eq36}
\end{equation}
The mass of the configuration is written as

\begin{equation}
 M_{3/2}=4\pi \left[\frac{5K}{8\pi G}\right]^{3/2}\rho_{ch}^{1/2}
{\hat M_{3/2}},\quad {\hat M_{3/2}}=\hat\rho_{0}^{1/2}\left[-\xi_{out}^2
\left(\frac{d\theta_{3/2}}{d\xi}\right)_
{out}+\frac{\beta\xi_{out}^3}{3}\right].
 \label{eq37}
\end{equation}
The density distribution for equilibrium configurations with
$\beta=0,\,\,\beta=0.5\beta_c,\,\,\beta=\beta_c$ is shown in Fig.5. The
nonphysical solution at $\beta=1.5\beta_c$, which has not an outer
boundary, is given by the dash-dot line.  At large $\xi$, these solutions
asymptotically approach the horizontal line $\theta_{3/2}=\beta^{2/3}$.
The numerical solution of equation (\ref{eq36}) gives $\beta_c=0.082$,
$\xi_{out}=3.654,\,\,3.984,\,\,5.086$, for $\beta=0,\,\,\beta=0.5\beta_c=
0.041,\,\,\beta=\beta_c=0.082$, respectively. In Fig.6 we show the
behavior of $\hat M_{3/2}(\hat\rho_0)|_\Lambda$, for different values of
$\beta_{in}=0,\,\,\beta_{in}=0.5\beta_{c},\,\,\beta_{in}=\beta_{c}$, for
which $\hat M_{3/2}=2.714,\,\,3.081,\,\,3.622$, at $\hat\rho_0=1$,
respectively.

\section{Application to the Local galaxy cluster}

The question about the importance of DE on the dynamics of the Local
Cluster (LC) was considered by Chernin (2001, 2008) and Chernin et
al. (2009). Simple estimations confirm this importance and, for the
presently accepted values of the DE density $\rho_v =
(0.72\pm 0.03)\cdot 10^{-29}$ g/cm$^3$, the mass of the local group,
including the dark mater input, is between $M_{LC} \sim 3.5\cdot
10^{12}\,M_\odot$, according to Chernin et al. (2009), and $M_{LC}
\sim 1.3\cdot 10^{12}\,M_\odot$, according to Karachentsev et al.
(2006). The radius $R_{LC}$ of the LC is known even worse. It may be
estimated by measuring the velocity dispersion $v_t$ of galaxies in
LC and by the application of the virial theorem, so that $R_{LC} \sim
\sqrt{(GM_{LC}/v_t})$. The velocity dispersion of the galaxies in LC,
estimated by Karachentsev et al. (2006), resulted equal to
$v_t= 63$ km/s, very close to the value of the local Hubble constant
$H=68$ km/s/Mpc (Karachentsev et al., 2006).

Similarity of this values indicates the great difficulties in dividing the
measured velocities between regular and chaotic components. In addition,
the indefiniteness increases because of the unknown level of the
anisotropy in the velocity distribution. So, the radius of the LC may be
estimated as $R_{LC}=(GM_{LC}/v_t^2)=(1.5\div 4)$ Mpc, with a very high
error box, which we cannot estimate properly. Chernin et al. (2009) give
different intervals for the mass and radius of LC, without using the fixed
value of $v_t$, but considering the whole picture of the distribution
of the galaxy velocities. They obtained

\begin{equation}
 1.2 < M_{LC} < 3.7 \cdot 10^{12}\,{\rm M_\odot} \quad {\rm and}
\quad 1.1 < R_v <1.6\,{\rm Mpc}.
 \label{eq41}
\end{equation}
It is important to note that Chernin et al. (2009) identified the
radius $R_{LC}$ of the LC with the radius $R_v$ of the zero-gravity force,
which is identical with the one corresponding to our critical model with
$\beta=\beta_c$, in which the average matter density is equal to
$2\rho_v$, as we can see from (\ref{eq19}). These estimations, while
having very big observational errors, indicate the importance of the
presently accepted value of DE density on the structure and dynamics
of the outer parts of LC, and its vicinity.

Polytropic solutions with DE are not well appropriate for describing the
LC, being the most mass concentrated in two giant objects: our Galaxy
and M31 (Andromeda). The polytropic model may have better application to
the rich galactic clusters, where the mass is more uniformly
distributed among a large number of galaxies and the distribution
of matter could be approximated by an averaged polytropic distribution.
In this aspect, it is necessary to make more extended measurements of
the galaxy velocities in the clusters and of their distribution over the
cluster extension.

\section{Conclusions}

The density of DE, measured from SN Ia distributions and spectra of CMB
fluctuations, imply the necessity to take into account such a contribution
in calculations of the structure of galaxy clusters. We developed these
calculations by considering simple polytropic models, in which it
is possible to see how the model changes with increasing of the influence
of DE.

Three different values of $n$ have been taken into account. Due to the
presence of DE, only for central densities larger than a critical value
$\rho_c$, depending of $n$, static solutions are found.  We have derived
the virial theorem for equilibrium configurations in presence of DE,
finding expressions between different sorts of the energy in
polytropic models.

 We have analyzed the stability of equilibrium configurations in presence
 of DE using energetic method. It is shown, that DE increases the
 instability of the equilibrium
configurations, working in the same direction as the influence of the
general relativistic corrections. The presence of DE decrease the the level
of relativity of a warm dark matter (WDM), at which it is still possible
to form gravitationally bound configurations from WDM.

Finally, the observational indefiniteness in the parameters of LC does not
permit to draw definite conclusions about the level of DE influence but,
without any doubt, it indicates the dynamic importance of DE in the scale
of the galaxy clusters.

\section*{Acknowledgments}

The part concerning the work of GSBK and SOT was partially supported by
the Russian Foundation for Basic Research grant 08-02-00491, the RAN
Program ``Origin, formation and evolution of objects of Universe" and
Russian Federation President Grant for Support of Leading Scientific
Schools NSh-3458.2010.2.

The authors are grateful to Dr. David Mota for useful comments.

\vfill\eject

\begin{figure}[h]
\centerline{\includegraphics[scale=0.5] {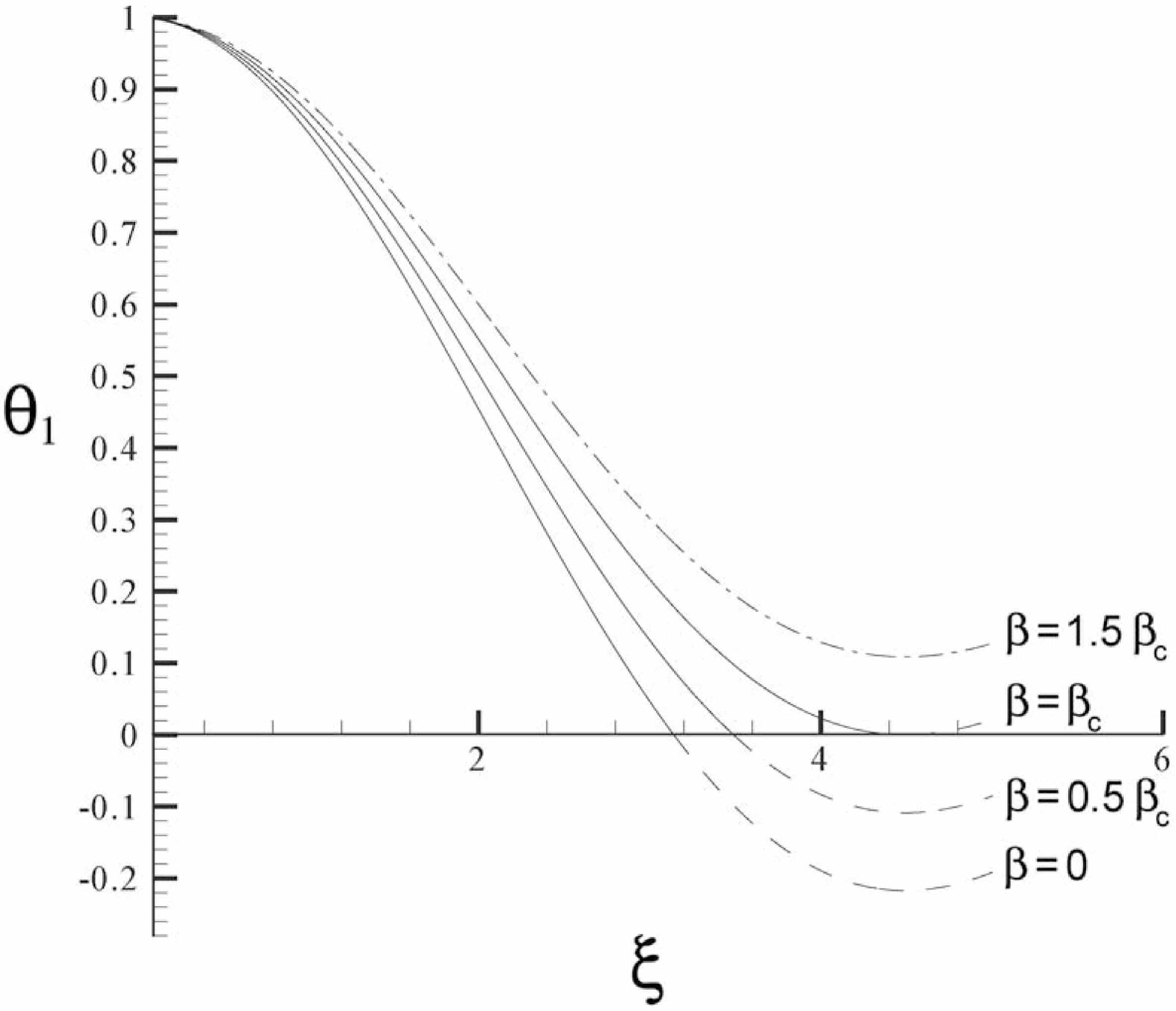} }
\caption{The density distribution for configurations at $n=1$ with
$\beta=0,\,\,\beta=0.5\beta_c,\,\,\beta=\beta_c$. The curves are marked
with the values of $\beta$. The nonphysical solution at
$\beta=1.5\beta_c$, which has not an outer boundary, is given by the
dash-dot line. The nonphysical parts of the solutions at $\beta\le
\beta_c$, behind the outer boundary, are given by the dash lines. The
solutions, at large $\xi$, asymptotically approach the horizontal line
$\theta_1=\beta$. \label{fig1}}
\end{figure}
\clearpage

\begin{figure}[h]
\centerline{\includegraphics[scale=0.5] {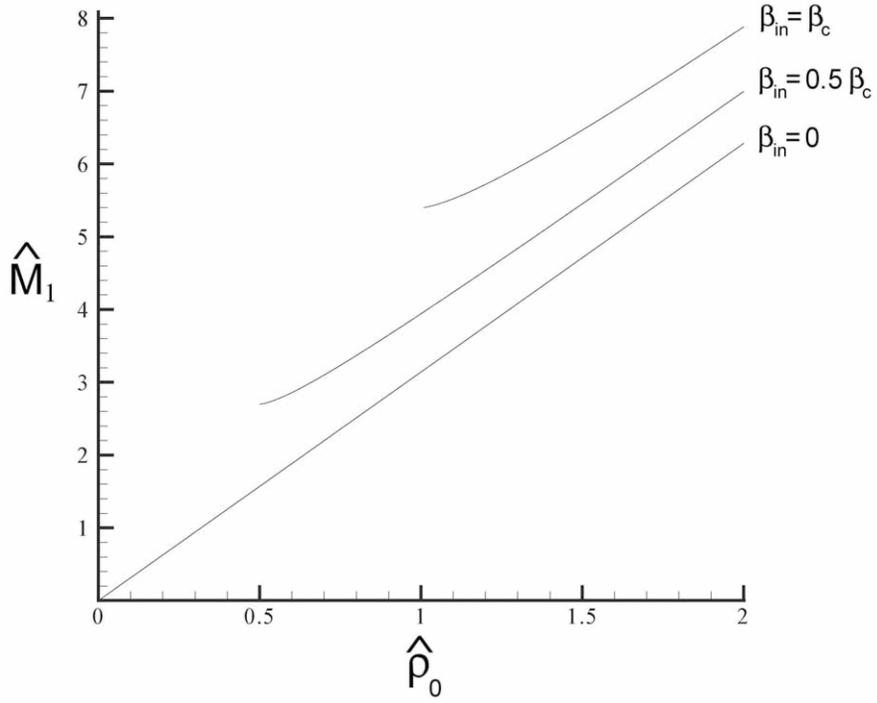} }
\caption{Nondimensional mass $\hat M_1$ of the equilibrium polytropic
configurations at $n=1$ as a function of the nondimensional central
density $\hat\rho_0$, for different values of $\beta_{in}$. The
cosmological constant $\Lambda$ is the same along each curve. The curves
at $\beta_{in}\neq 0$ are limited by the configuration with
$\beta=\beta_c$. \label{fig2}}
\end{figure}
\clearpage

\begin{figure}[h]
\centerline{\includegraphics[scale=0.5] {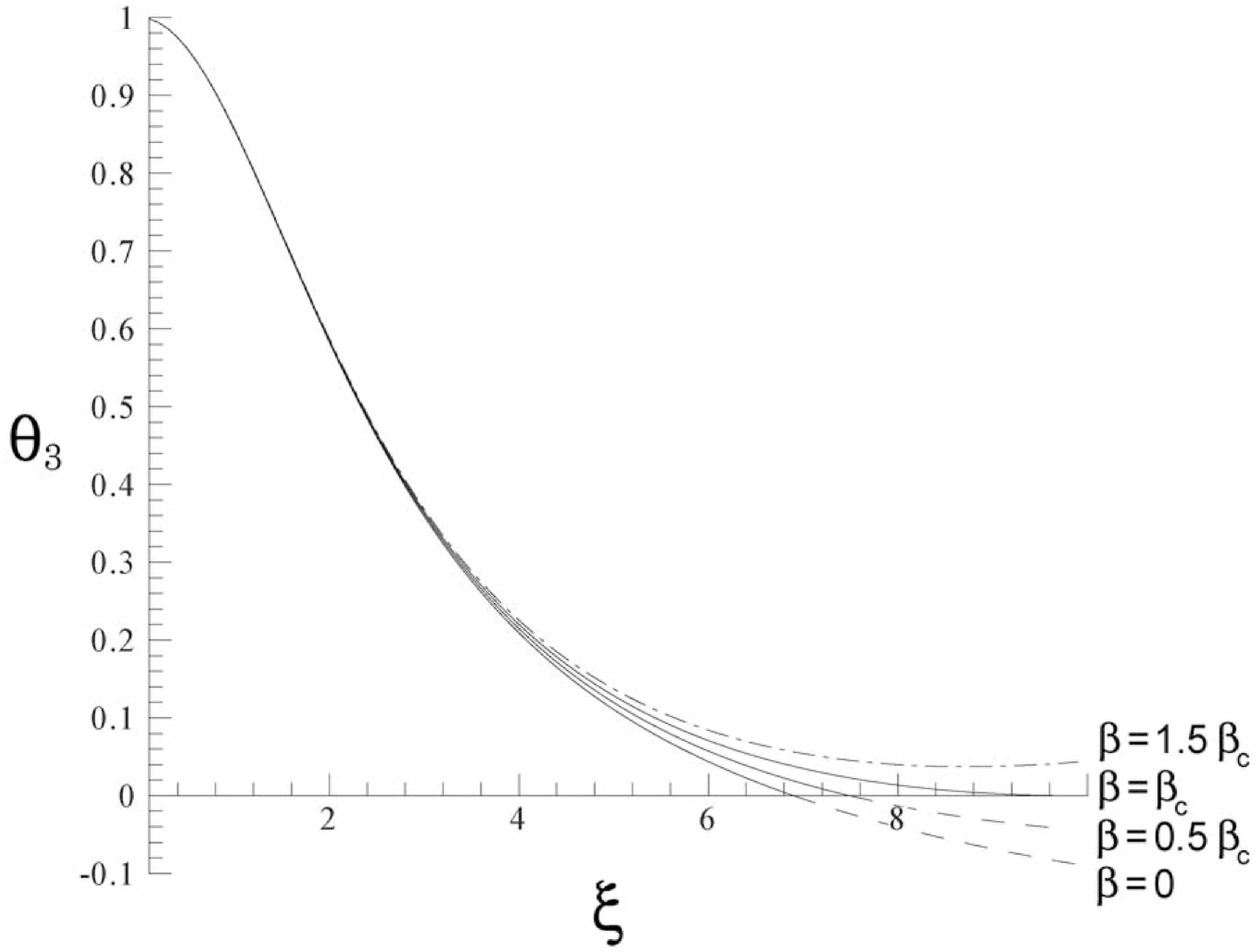} }
\caption{Same as in Fig.1, for $n=3$. The solutions asymptotically
approach, at large $\xi$, the horizontal line $\theta_3=\beta^{1/3}$.
\label{fig3}}
\end{figure}
\clearpage

\begin{figure}[h]
\centerline{\includegraphics[scale=0.5] {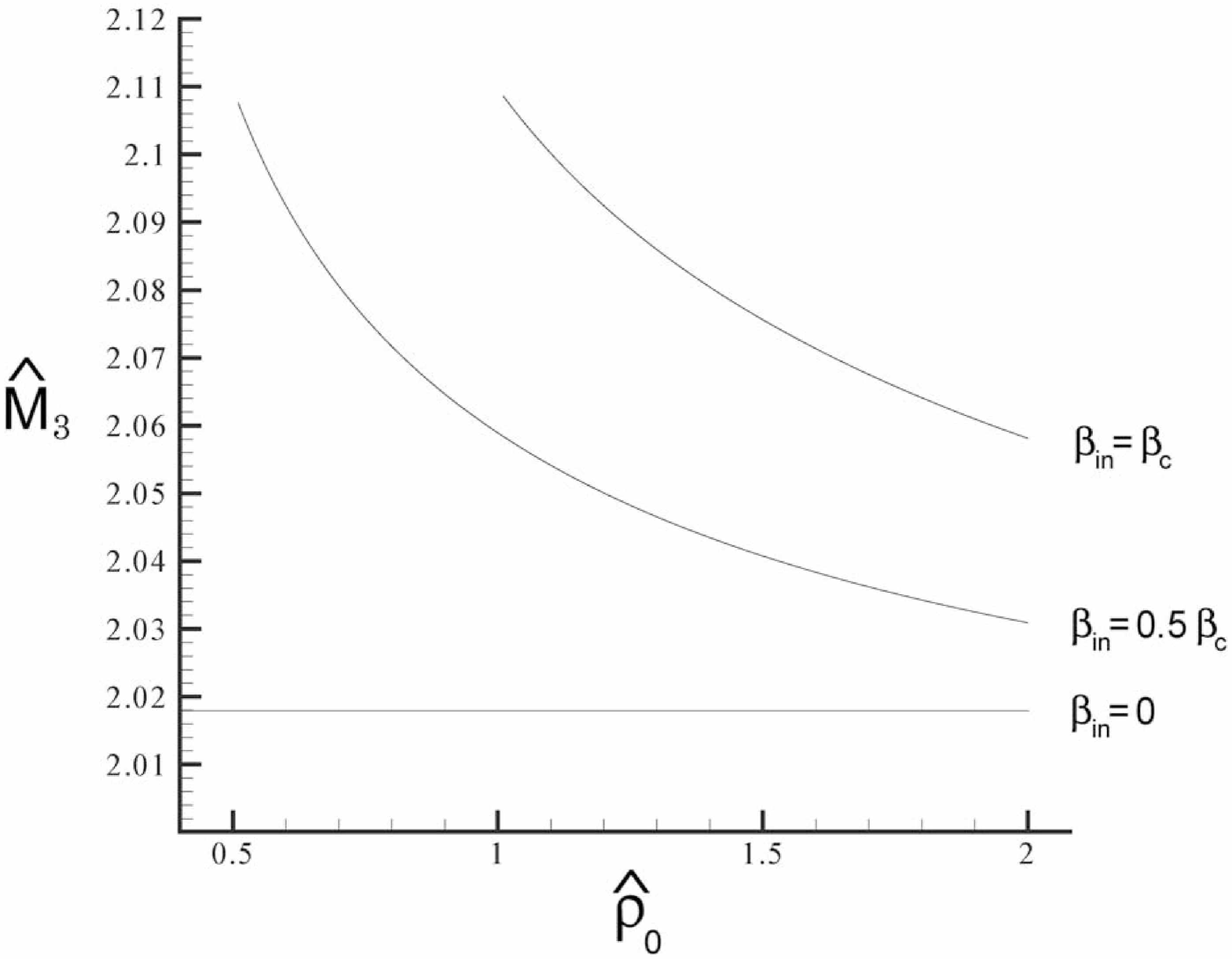} }
\caption{Same as in Fig.2, for $n=3$. \label{fig4}}
\end{figure}
\clearpage

\begin{figure}[h]
\centerline{\includegraphics[scale=0.5] {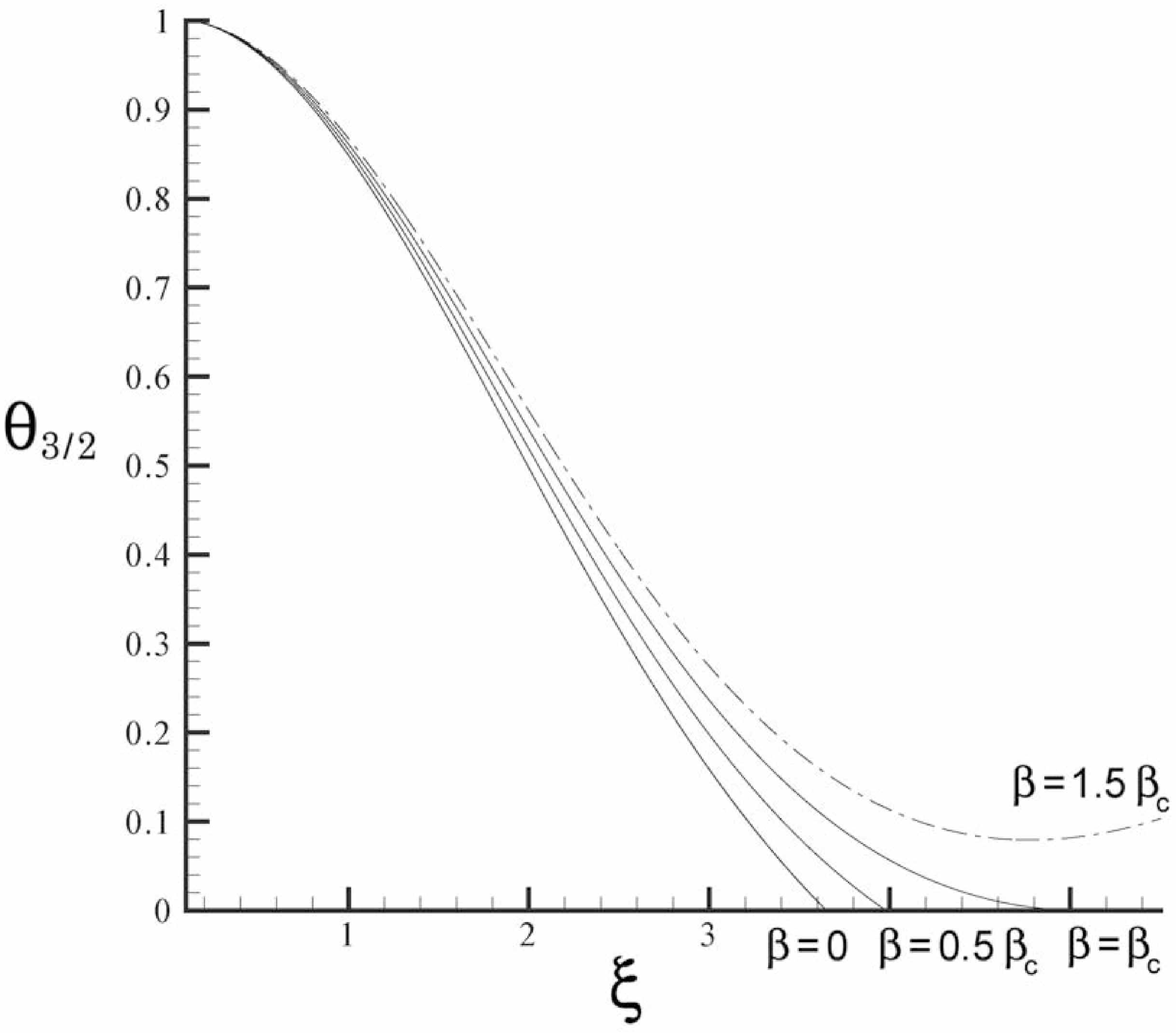} }
\caption{Same as in Fig.1, for $n=3/2$. The solutions asymptotically
approach, at large $\xi$, the horizontal line $\theta_{3/2}=\beta^{2/3}$.
\label{fig5}}
\end{figure}
\clearpage

\begin{figure}[h]
\centerline{\includegraphics[scale=0.5] {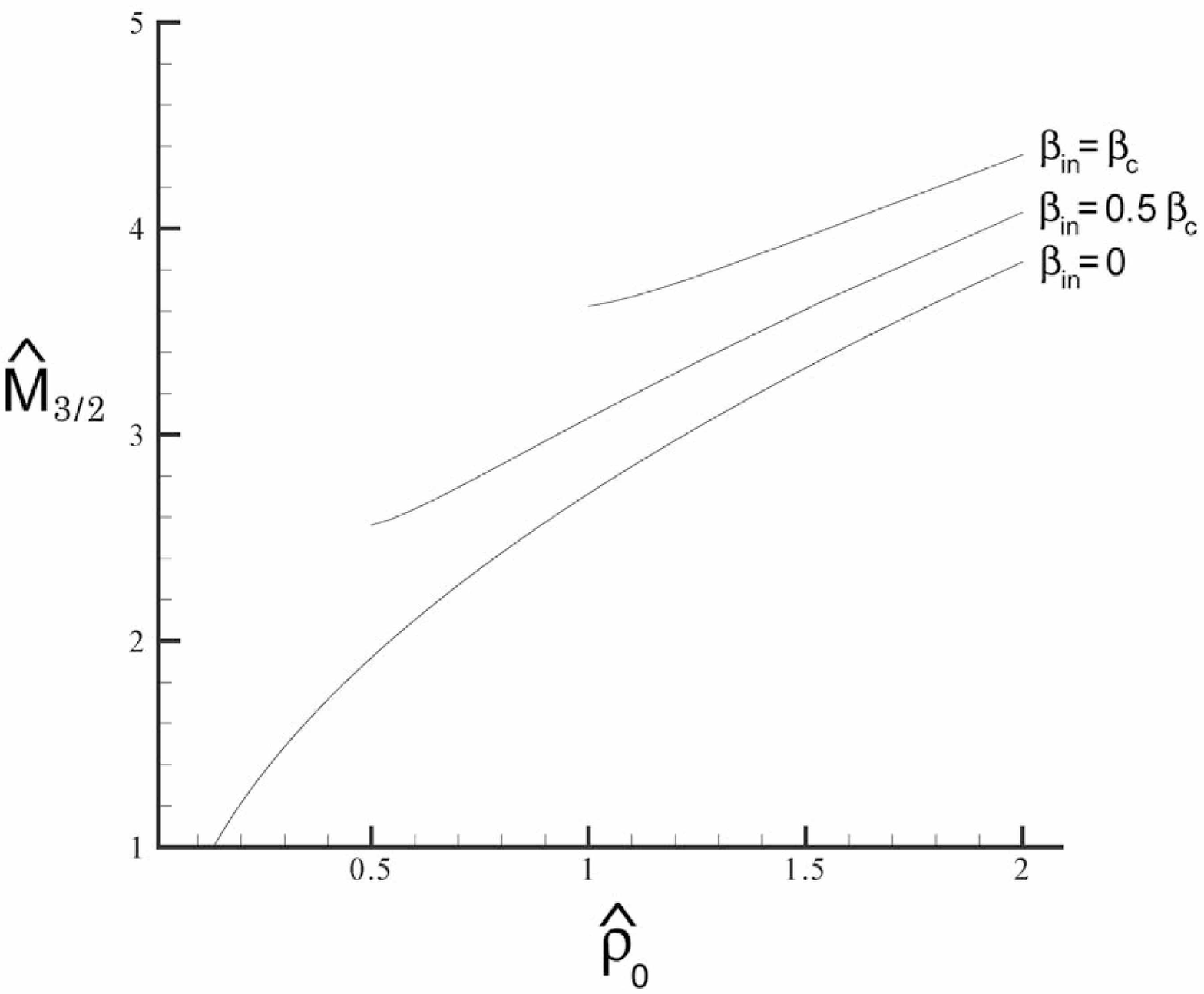} }
\caption{Same as in Fig.2, for $n=3/2$. \label{fig6}}
\end{figure}
\clearpage

\end{document}